\documentclass[epj,final]{svjour}
\usepackage{latexsym}
\usepackage{graphicx}
\usepackage{amsmath}
\usepackage{amsfonts}
\usepackage{amssymb}

\begin{document}

\title{Hadronic shift in pionic hydrogen}

\author{M. Hennebach\inst{1}\thanks{\emph{present address:} 
        Areva GmbH, D-63067 Offenbach, Germany},
        D.\,F.\,Anagnostopoulos\inst{2},
        A.\,Dax\inst{3}, 
        H.\,Fuhrmann\inst{4}, 
        D.\,Gotta\inst{1}\thanks{\emph{corresponding author:} d.gotta@fz-juelich.de}, 
        A.\,Gruber\inst{4}\thanks{\emph{present address:}
        Universit\"atsklinik f\"ur Radiologie und Nuklearmedizin, Medizinische Universit\"at Wien, 
        A-1090 Vienna, Austria},
        A.\,Hirtl\inst{4c}, 
        P.\,Indelicato\inst{5,6,7},
        Y.-W.\,Liu\inst{3}\thanks{\emph{present address:} Physics Department, National Tsing Hua University, Hsinchu 300, Taiwan},
        B.\,Manil\inst{5}\thanks{\emph{present address:}
        Laboratoire de Physique des Lasers, CNRS et Universit\'e Paris 13, France},
        V.\,E.\,Markushin\inst{3}\thanks{\emph{present address:} Department of IT, Paul Scherrer Institut, CH-5232 Villigen, Switzerland},
        A.\,J.\,Rusi\,el\,Hassani\inst{8},
        L.\,M. Simons\inst{3},
        M.\,Trassinelli\inst{9,10},
\and 
        J.\,Zmeskal\inst{4}
        }

\institute
{
Institut f\"ur Kernphysik, Forschungszentrum J\"ulich, D-52425 J\"ulich, Germany
\and 
Department of Materials Science and Engineering, University of Ioannina, GR-45110 Ioannina, Greece 
\and 
Laboratory for Particle Physics, Paul Scherrer Institut, CH-5232 Villigen, Switzerland
\and 
Stefan Meyer Institute for Subatomic Physics, Austrian Academy of Sciences, A-1090 Vienna, Austria 
\and 
Laboratoire Kastler Brossel,  Sorbonne Universit\'es, UPMC Univ. Paris 06, Case 74; 4, place Jussieu, 75005 Paris, France
\and
Laboratoire Kastler Brossel,  CNRS, 75005, Paris, France
\and
Laboratoire Kastler Brossel, D\'epartement de Physique de l'\'Ecole Normale Sup\'erieure,  24 Rue Lhomond, 75005, Paris, France
\and 
Facult\'e des Sciences et Techniques, Universit\'e Abdelmalek Essa\^adi, Tanger, Morocco
\and
Institut des NanoSciences de Paris, Sorbonne Universit\'es, UPMC Univ. Paris 06, Case 840; 4, place Jussieu, France
\and
Institut des NanoSciences de Paris,  CNRS, F-75005, Paris, France
}
           
\date{Received: date / Revised version: date}
\abstract{
The hadronic shift in pionic hydrogen has been redetermined to be $\epsilon_{1s}=7.086\,\pm\,0.007(stat)\,\pm\,0.006(sys)$\,eV by X-ray spectroscopy of ground state transitions applying various energy calibration schemes. The experiment was performed at the high-intensity low-energy pion beam of the Paul Scherrer Institut by using the cyclotron trap and an ultimate-resolution Bragg spectrometer with bent crystals.
\PACS{{36.10.Gv, 25.80.Ls, 07.85.Nc}{Mesonic atoms, Pion-nucleon scattering, X-ray spectrometers}} 
}

\authorrunning{M.\,Hennebach {\it et al.}}
\titlerunning{Hadronic shift in pionic hydrogen}

\maketitle

\section{Introduction}
In hadronic atoms, the strong pion-nucleus interaction manifests itself by a change of binding energies and level widths of the low-lying atomic levels. It can be measured in the energies and line widths of the corresponding X-ray transitions when compared to a pure electromagnetically bound system. 

Experimentally accessible in the lightest pionic atom ---pionic hydrogen ($\pi$H)--- are the X-ray transitions to the 1s ground state, which terminate the atomic de-excitation cascade. Any observed influence of the strong interaction can be attributed fully to the 1s state, because hadronic and collisional effects in $p$ states are negligibly small. Hence, the $s$-level shift and width in $\pi$H directly measure the $\pi^{-}p$ $s$-wave interaction.

The hadronic level shift in pionic hydrogen is related to the $\pi^{-}p$ scattering lengths in leading order by the Deser-Goldberger-Baumann-Thirring (DGBT) formula\,\cite{Des54}
\begin{eqnarray}
\frac{\it\epsilon_{1s}}{B_{1s}} &=& -~\frac{4}{r_{B}}\,\,a^{LO}_{\pi^{-}p}\,\label{eq:deser}.
\end{eqnarray} 
The quantity $a^{LO}_{\pi^{-}p}$ denotes the scattering length, $B_{1s}=-\alpha^{2}\mu c^{2}/2$ is the Coulomb point-nucleus ground-state binding energy, $r_B=\hbar c/\alpha\mu c^{2}$ the Bohr radius of the pionic hydrogen atom, and $\mu=m_{\pi}/(1+\frac{m_{\pi}}{m_p})$ the reduced mass of the $\pi$H system. 

The correction due to the fact, that the $\pi$H system is a Coulomb bound state, usually is taken into account by Trueman's expansion in the ratio of scattering length to the Bohr radius\,\cite{Tru61,Lam69,Lam70,Mit01}. In the case of hydrogen ($Z=1$) it can be expressed up to order $O(\alpha^4)$ and including the logarithmic term as follows\,\cite{Lyu00} 
\begin{eqnarray}
\epsilon_{1s}&=& \frac{2\alpha^{3}\mu^{2}c^{4}}{\hbar c}\,a_{\pi^-\mathrm{p}}\label{eq:eps_moddeser}\\
             & &\cdot\,\left[1-\frac{2\alpha\mu c^2}{\hbar c}\,(\mathrm{ln}\,\alpha -1)
                \cdot a_{\pi^-\mathrm{p}}+\delta^{\mathrm{vac}}_{\epsilon}\right].\nonumber
\end{eqnarray}
The term $\delta^{\mathrm{vac}}_{\epsilon}$\,=\,0.48\% accounts for the interference of vacuum polarisation and strong interaction\,\cite{Eir00}. Its uncertainty is assumed to be negligibly small compared to the experimental accuracy\,\cite{Gas02}. The quantity $a_{\pi^{-}p}$ then represents the  scattering length for two strongly interacting charged particles.

The derivation of the pure hadronic part of scattering lengths is an ongoing effort spanning several decades, both with phenomenological and microscopic approaches. The work is summarized in a recent review, which emphasizes mainly the ansatz of an effective field theory as Chiral Perturbation Theory ($\chi$PT)\,\cite{Gas08}. Recent phenomenological descriptions are given in refs.\,\cite{Eri04,Mat12,Mat13}. 

Within the framework of $\chi$PT, the extraction of pure QCD quantities is based on the treatment of strong and electromagnetic isospin violating effects at the same level. At threshold, the two basic parameters of the $\pi N$ interaction are the isoscalar and isovector $s$-wave scattering lengths $a^{+}$ and $a^{-}$. In the limit of isospin conservation and the absence of electric charges, they are given in terms of the elastic reactions $\pi^{-}p\to\pi^{-}p$ and $\pi^{+}p\to\pi^{+}p$ by
\begin{eqnarray}
a^{\pm} &\equiv& \frac{1}{2}\,\,(a_{\pi^{-}p}\pm a_{\pi^{+}p})\,,\label{eq:apm}
\end{eqnarray} 
{\it i.\,e.}, $a_{\pi^{-}p} = a^{+}+a^{-}$ holds.

The corrections due to strong ($m_u\neq m_d$) and electromagnetic ($Q=e$) isospin violation have been calculated at various levels and approaches \cite{Lyu00,Gas02,Ber95}. Numerical values are based on preliminary results for $\epsilon_{1s}$ and $\Gamma_{1s}$ of this experiment\,\cite{Got08}. Within the framework of $\chi$PT, the most recent determination worked out in next-to-leading order (NLO) is given in refs.\,\cite{Hof09,Hof10}. The scattering length can be expressed in the form\,\cite{Hof09,Hof10,Bar11a,Bar11b} 
\begin{eqnarray}
a_{\pi^{-}p} = (a^{+}+a^{-})+\bigtriangleup a_{\pi^{-}p}\,,\label{eq:api-p_ChiPT}
\end{eqnarray} 
where the numerical value of the correction is $\bigtriangleup a_{\pi^{-}p}\approx(-7.7\,\pm\,3.3)\cdot 10^{-3}\,m^{-1}_{\pi}$. The uncertainty is dominated ($\pm\,3.0\cdot 10^{-3}\,m^{-1}_{\pi}$) by the poor knowledge of the low-energy constants (LECs) $f_1$ and $c_1$ appearing in leading order of the chiral expansion.

In order to extract the individual scattering length $a^{+}$ or $a^{-}$, the information obtained from $\epsilon_{1s}$ must be combined with the  ground-state broadening $\Gamma_{1s}$, which depends in leading order only on $a^{-}$. In a similar form as given in eq.\,(\ref{eq:eps_moddeser}), $\Gamma_{1s}$ reads\,\cite{Zem04}   
\begin{eqnarray}
\Gamma_{1s}&=& \frac{4\alpha^{3}\mu^{2}c^{4}}{\hbar c}\,q_0\,
                \left(1+\frac{1}{P}\right)(a^{\mathrm{cex}}_{\pi^{-}p})^2\label{eq:Ga_moddeser}\\
           & & \cdot\,\left[1-\frac{4\alpha\mu c^{2}}{\hbar c}\,(\mathrm{ln}\,\alpha -1)\cdot a_{\pi^{-}p}\right.\nonumber\\
           & & \left.+\frac{2\mu c^{4}}{(\hbar c)^{2}}(m_{p}+m_{\pi^{\pm}}-m_{n}-m_{\pi^{0}})\cdot a^{2}_{\pi^{0}n}+\delta^{\mathrm{vac}}_{\epsilon}\right].\nonumber
\end{eqnarray}
Here, $q_0$ denotes the center-of-mass momentum of the $\pi^0$ and $P$ for pion capture at rest the Panofsky ratio measured to be $P=\frac{\sigma(\pi^-p\rightarrow \pi^0n)}{\sigma(\pi^-p\rightarrow \gamma n)}=1.546\,\pm\,0.009$\,\cite{Spu77}. The scattering length $a_{\pi^{0}n}$ is well approximated by $a^+$\,\cite{Bar11b}.

The QCD quantity $a^{-}$ is obtained by applying appropriate corrections to the scattering length $a^{\mathrm{cex}}_{\pi^{-}p}$ of the charge exchange reaction $\pi^{-}p\rightarrow\pi^{0}n$:  
\begin{eqnarray}
a^{\mathrm{cex}}_{\pi^{-}p} = -\sqrt{2}\,a^{-}+\bigtriangleup a^{\mathrm{cex}}_{\pi^{-}p}\,.\label{eq:api-p_cex_ChiPT}
\end{eqnarray} 
Within the framework of $\chi$PT, one obtains $\bigtriangleup a^{\mathrm{cex}}_{\pi^{-}p}=(0.4\,\pm\,0.9)\cdot 10^{-3}\,m^{-1}_{\pi}$\,\cite{Hof09,Hof10}. The significantly smaller uncertainty is due to the fact that in this case the LECs $f_1$ and $c_1$ do not appear in leading order.

Further information on the scattering lengths is obtained from the ground-state shift $\epsilon^{\pi\mathrm{D}}_{1s}$ in pionic deuterium. It is related to the real part of the $\pi$D scattering length constituting in leading order the coherent sum of $\pi^{-}p$ and $\pi^{-}n$ scattering, which is given by $2\cdot a^+$.  Consequently, the triple  $\epsilon^{\pi\mathrm{H}}_{1s}$, $\Gamma^{\pi\mathrm{H}}_{1s}$, and $\epsilon^{\pi\mathrm{D}}_{1s}$ provides a constraint on the two scattering lengths $a^{+}$ and $a^{-}$. The theoretical framework may be found in refs.\,\cite{Bar11a,Bar11b}.

Ultimate precision X-ray spectroscopy of pionic and muonic atoms became feasible with the advent of meson factories allowing an efficient use of crystal spectrometers\,\cite{Got04}. The measurements described here are part of a series of experiments\,\cite{PSI98} aiming at a new precision determination of $\epsilon^{\pi\mathrm{H}}_{1s}$, $\Gamma^{\pi\mathrm{H}}_{1s}$, and $\epsilon^{\pi\mathrm{D}}_{1s}$ in order to allow for a decisive constraint on $a^{+}$ and $a^{-}$. In particular, a comprehensive study of the hadronic line broadening was performed to investigate cascade effects affecting the line width. Therefore, also as an independent check, a study of the cascade-induced line broadening was performed in the twin-system muonic hydrogen\,\cite{Cov09}. 

In this experiment, three different transitions to the ground state  ---$\pi$H$(2p-1s)$, $\pi$H$(3p-1s)$, and $\pi$H$(4p-1s)$ (K$\alpha$, K$\beta$, and K$\gamma$)--- have been measured. For energy calibration, both fluorescence and  ---for the first time--- X-rays from exotic atoms were used. The hydrogen density varied from 4\,bar to liquid (LH$_2$) in order to study the influence of cascade effects both on X-ray energy and line shape. 

This paper reports values for $\epsilon^{\pi\mathrm{H}}_{1s}$ from those measurements for which an energy calibration is available with sufficient accuracy. Results for $\epsilon^{\pi\mathrm{D}}_{1s}$ and $\Gamma^{\pi\mathrm{D}}_{1s}$ are given in detail in refs.\,\cite{Str10,Str11}. Final results of the measurement of $\Gamma^{\pi\mathrm{H}}_{1s}$ will be discussed elsewhere\,\cite{Hir14}. Preliminary values are given in ref.\,\cite{Got08}.

\section{Atomic cascade}\label{sec:cascade}

\begin{center}
\begin{figure}[t]
\resizebox{0.48\textwidth}{!}{\includegraphics{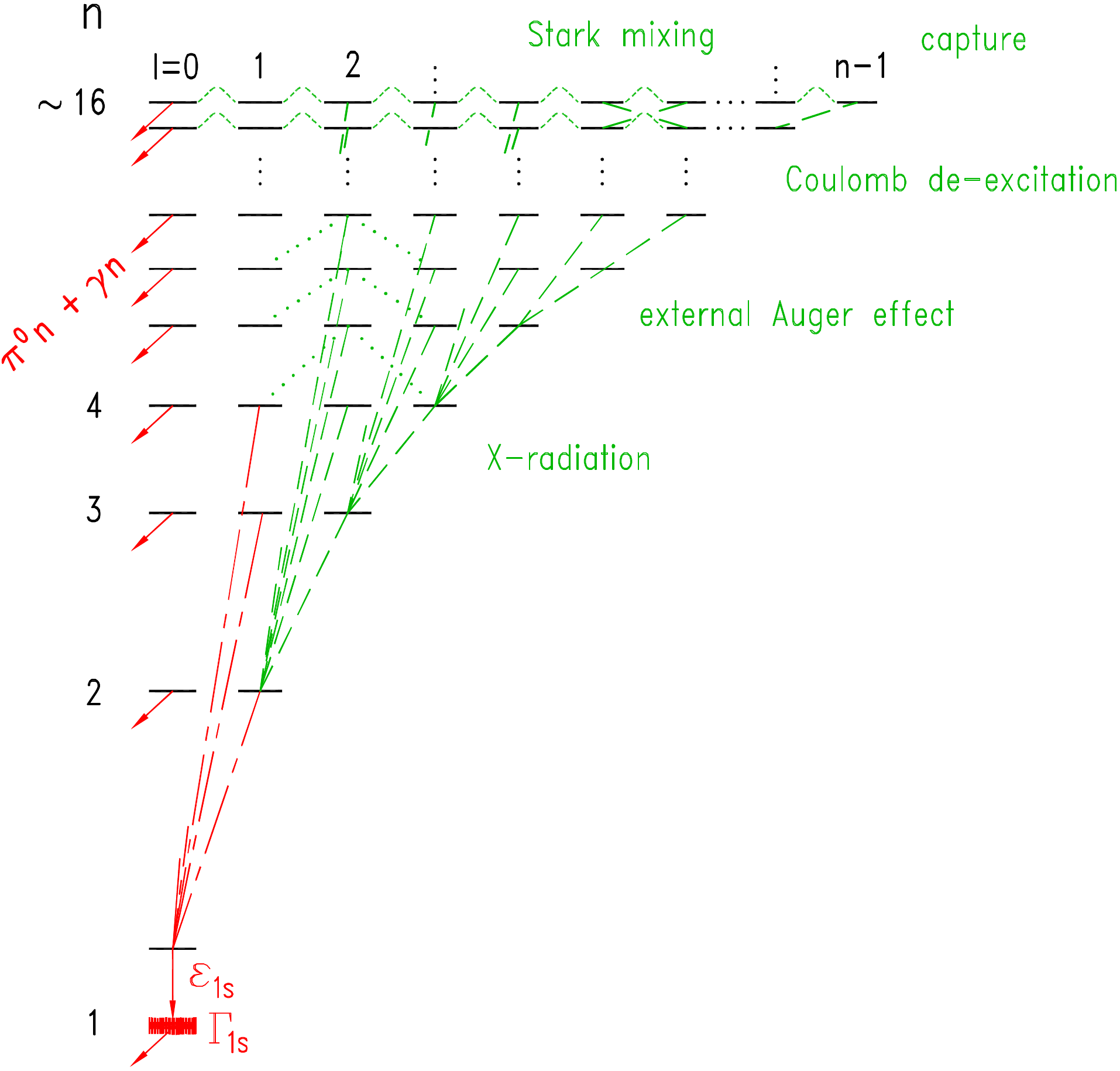}}
\caption{Atomic de-excitation cascade in pionic hydrogen. Collisional processes and X-radiation dominate the upper and the lower  part of the atomic cascade, respectively. X-ray energies of the 3 most intense transitions $\pi$H$(2p-1s)$, $\pi$H$(3p-1s)$, and $\pi$H$(4p-1s)$ are 2.437, 2.885 and  3.043\,keV. Hadronic shift $\epsilon_{1s}$ and width $\Gamma_{1s}$ are on the order of +\,7 and 1\,eV, respectively. The hadronic shift is defined as $\epsilon_{1s}\equiv \mathrm{E_{exp}-E_{QED}}$, {\it i.\,e.}, a positive sign, as is the case for $\pi$H, corresponds to an attractive interaction.}
\label{fig:fig1_cascad_pih}
\end{figure}
\end{center}

After capture of the pion into atomic states, a quantum cascade starts approximately at principle numbers $n_{\pi}=16$\,\cite{Coh04}. During this cascade the  pionic-hydrogen atom undergoes many collisions with H$_{2}$ molecules, where de-ex\-ci\-ta\-tion occurs mainly by (external) Auger electron emission or Coulomb de-excitation. The lower part of the cascade is dominated more and more by X-radiation (fig.\,\ref{fig:fig1_cascad_pih}). 

In Auger emission, the released energy is almost completely transferred to the electron because of the mass ratio. In Coulomb de-excitation the $\pi$H system encounters a proton in a  molecule and the de-excitation energy is shared about half and half between the collision partners leading to significant acceleration. The latter process leads to a Doppler broadening which is symmetric and, hence, does not affect the transition energies.

The electrically neutral $\pi$H system, being small compared to atomic dimensions, easily penetrates the Coulomb field of the target atoms where the electric field induces transitions between angular momentum states for constant $n_{\pi}$ (Stark mixing). Due to the strong interaction the induced $s$-level width causes the removal of pions from the cascade by the charge exchange $\pi^{-}p\rightarrow \pi^{0}n$ and radiative capture reaction $\pi^{-}p\rightarrow \gamma n$. This leads to a drastic reduction of the X-ray line yields with increasing target density (Day-Snow-Sucher effect).

Measured line yields for K$\alpha$, K$\beta$, and K$\gamma$ lines have been reported for equivalent densities between 3 and 40\,bar to be on the order of a few percent\,\cite{Has95}. A detailed discussion of the atomic  cascade in exotic hydrogen may be found in refs.\,\cite{Har90,Jen03}.

In order to determine the hadronic shift $\epsilon^{\pi\mathrm{H}}_{1s}$ precisely, collision induced  processes causing line shifts must be considered:

(i) Molecular formation of metastable hybrid molecules $[(pp\mu )p]ee$ during $\mu $H + H$_2$ collisions is well known from the study of muon-catalyzed fusion and muonic hydrogen even at lowest densities and in excited states\,\cite{Die13}. During collisions, an analogous process $\pi^{-} p+\mathrm{H}_{2}\rightarrow [(pp\pi^{-})p]ee$ must be expected  \,\cite{Jon99}. Such complex molecules are assumed to stabilise non-radiatively by Auger emission\,\cite{Kil04}. 

Possible X-ray transitions from weakly bound molecular states before stabilisation  falsify the value for the hadronic shift determined from the measured X-ray energy due to satellite lines shifted to lower energies. In contrast, Auger stabilised molecules emit X-rays of at least 30\,eV energy less than the undisturbed transition and are easily resolved in this experiment. 

As molecular formation is collision induced, the fraction of formed complex molecules and the X-ray rate would depend on the target density. X-ray transitions from molecular states would show up as low-energy satellites with density dependent intensities because of different  collision probabilities. Though radiative decay is predicted to be small in the case of $\pi$H\,\cite{Kil04,Lin03}, a measurement of the $\pi$H$(3p-1s)$ X-ray energy was performed at different densities to possibly identify such radiative contributions.

(ii) The induced energy shift due to Stark broadening has been estimated to be maximal -\,4\,meV at liquid hydrogen density\,\cite{Jen03a}. Consequently, this contribution is neglected at all  lower densities.

(iii) Estimates on the pressure shift of the X-ray energy yield values less than $-0.2$\,meV at all densities considered.

\section{Experimental approach}

\begin{figure}[t]
\resizebox{0.49\textwidth}{!}
{\includegraphics{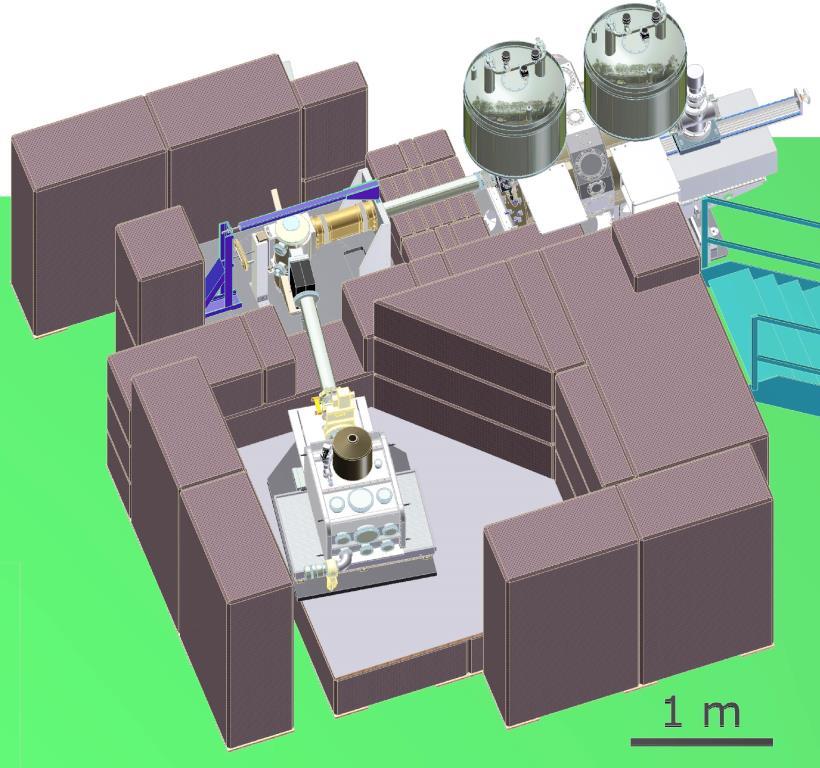}}
\caption{Set-up of cyclotron trap (top right) and crystal spectrometer in the $\pi$E5 area at PSI for the $\pi$H$(3p-1s)$ (quartz $10\bar{1}$) and the $\pi$H$(4p-1s)$ transition (silicon 111). The roof of the concrete cave is not shown.}
\label{fig:setup}
\end{figure}

\begin{figure}[t]
\resizebox{0.49\textwidth}{!}
{\includegraphics{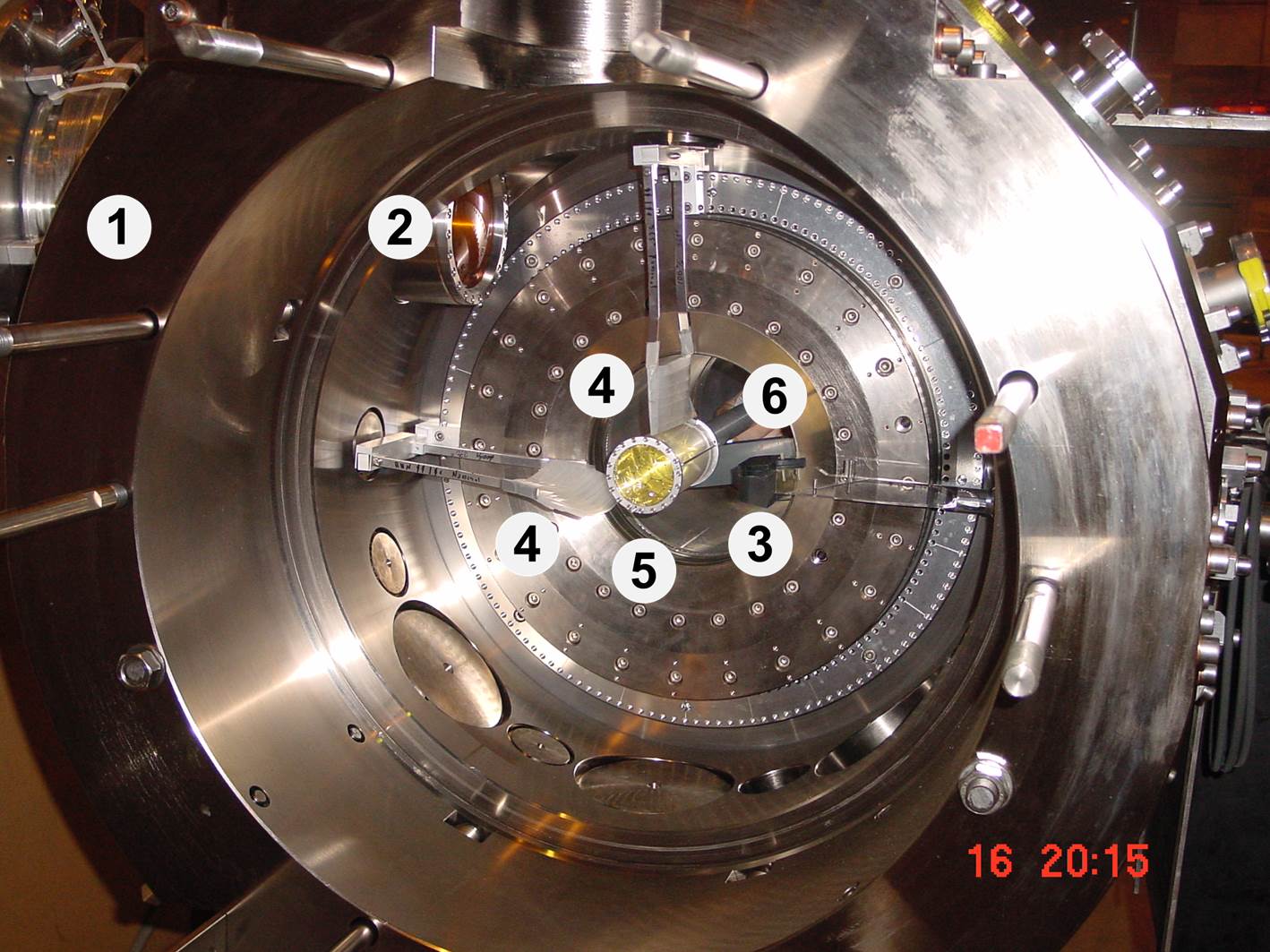}}
\caption{Installations inside the cyclotron trap. One coil of the magnet assembly is removed. The inner part of the magnetic shielding (iron return yoke) serves as wall for the vacuum chamber containing target cell and degraders. 1) return yoke, 2) beam entrance window, 3) first  degrader (38\,mm carbonized polyethylene), 4) wedge-shaped secondary degraders (stack of polyethylene foils), 5) gas cell, 6) cooling finger.}
\label{fig:cyclotron_trap}
\end{figure}

The experiment used the high intensity pion beam extracted to the $\pi$E5 area at the Paul Scherrer Institute (fig.\,\ref{fig:setup}). The average proton current of the accelerator at the pion-production target was between 1.35 and 1.8\,mA. The pion beam line was set up to 85\,MeV/c momentum and guided into  the cyclotron trap II\,\cite{Sim88}. In the center of the trap a gas-filled cylindrical target cell was placed (fig.\,\ref{fig:cyclotron_trap}). A fraction of about 0.5\% of the incoming pions was stopped in hydrogen gas per 1\,bar equivalent pressure. The cell wall was made of a 75\,$\mathrm{\mu}$m thick Kapton foil with a diameter of about 59\,mm. Its end cap towards the crystal spectrometer contained a 7.5\,$\mathrm{\mu}$m Kapton window supported either by a stainless steel honeycomb structure or by narrow horizontally oriented aluminum bars. 

As a thin Kapton window limits gas pressure to about 2\,bar, the hydrogen density was adjusted by means of a cooling finger by temperature variation. In this way, a density range from 4\,bar equivalent to liquid (corresponding to 700\,bar) was covered. The  density of at least 4\,bar is necessary in order to achieve a sufficient count rate.

The free length of the target cell's Kapton wall was about 120\,mm except for the case $\pi$H$(4p-1s)$/Ga K$\alpha_2$. Here, a cell of  220\,mm length was used with a GaAs plate installed inside the gas volume as used for the experiment described in ref.\,\cite{Str11}. In the case of the combination $\pi$H$(3p-1s)$/$\pi$Be$(4f-3d)$, thin beryllium plates were installed inside the target cylinder. The arrangement was similar to the one outlined in ref.\,\cite{Sig96a} where the $\pi$Be line was used to determine the spectrometer response. 

The crystal spectrometer was set up in Johann geo\-metry\,\cite{Joh31}. In this experiment, spherically bent Bragg crystals were used resulting in a partial vertical focussing which increases the count rate. Such a configuration allows the simultaneous measurement of a complete energy interval as given by the extension of the target in the direction of dispersion and when using a detector of corresponding size. In this way, the simultaneous measurement became possible for the line pairs $\pi$H$(3p-1s)$/$\pi^{16}$O$(6h-5g)$ and $\pi$H$(3p-1s)$/$\pi$Be$(4f-3d)$ (see table\,\ref{table:Bragg_angles}).

For X-ray diffraction, a quartz and a silicon crystal was used cut along the $(10\bar{1})$ and $(111)$ plane, respectively. Angular resolutions\,\cite{Ana05}, a possible miscut and its orientation\,\cite{Cov08} were determined in dedicated experiments. The bending radii of the quartz and the silicon crystal were measured to $R_c=(2980.6\,\pm\,0.4$) and $(2982.2\,\pm\,0.3)$\,mm\,\cite{ZEISS}. The full diameter of the crystal plates of  100\,mm was limited twofold: (i) by a spherical aperture to a diameter of  95\,mm in order to avoid edge effects and (ii) in the direction of dispersion to $\pm\,30$\,mm to keep Johann broadening small\,\cite{Egg65}.

A $3\times 2$ array ($vertical\times horizontal$) of charge-coupled devices having a sensitive area of $72\times 48$\,mm$^{2}$ was used for position-sensitive X-ray detection in two dimensions\,\cite{Nel02,Ind06} as in the muonic hydrogen\,\cite{Cov09} and pionic deuterium experiments\,\cite{Str10,Str11}. The detector's distance from the\linebreak
Bragg crystal is found by the focussing condition\linebreak
$R_c\cdot\sin\,\theta_{\mathrm{Bragg}}$ corrected for the miscut\,\cite{Cov08}. 

The spectrometer was set up inside a massive concrete cave (fig.\,\ref{fig:setup}). In this way, beam induced background, in particular from neutrons, is substantially suppressed. The cave serves, in addition, as (passive) temperature stabilisation. Thus, the temperature was kept stable to about $\pm\,1^{\circ}$ damping the variation in the environment by one order of magnitude. Thermal expansion coefficients of quartz and silicon are $14\cdot10^{-6}$ and  $2.5\cdot10^{-6}$ so that changes of the crystals' lattice constants cause only negligibly small shifts. Details of the interconnections of the chamber housing the Bragg crystal to the cyclotron trap containing the target cell on one side and the cryostat with the X-ray detector are displayed in figs.\,4 and 5 of ref.\,\cite{Str11}. 

An accurate determination of the X-ray energy requires both (i) sufficient statistics and (ii) the possibility to determine the line's center-of-gravity very precisely. For the $\pi$H line a sufficiently high X-ray rate is not achievable at low densities. Because of Stark mixing, liquid hydrogen is also discarded. The best figure of merit is obtained for an equivalent target density between 10 and 30\,bar (table\,\ref{table:exp_conditions}). Therefore, 30\,bar was chosen for a high statistics measurement of the $\pi$H$(3p-1s)$ transition, which spanned about 3 weeks. 

During the long-term measurements (labels C and D in table\,\ref{table:exp_conditions}) a mechanical shift of the X-ray detector was observed. It resulted in a monotonic position shift of the $\pi$H$(3p-1s)$ line. The shift of the $\pi$O$(6h-5g)$ calibration line was found to occur synchronously. The line shift was corrected by means of a polynomial ansatz. The uncertainty of the correction ($\pm\,4.5$\,meV) is the largest contribution to the systematic errors of measurements B and D (table\,\ref{table:results_R98}). 

A comprehensive discussion can be found in ref.\,\cite{Hen03}. The mechanical setup is similar to the one used in the pionic-deuterium-experiment and is described in detail in ref.\,\cite{Str11}.

\section{Energy calibration}

\begin{figure}[b]
\resizebox{0.42\textwidth}{!}{\includegraphics{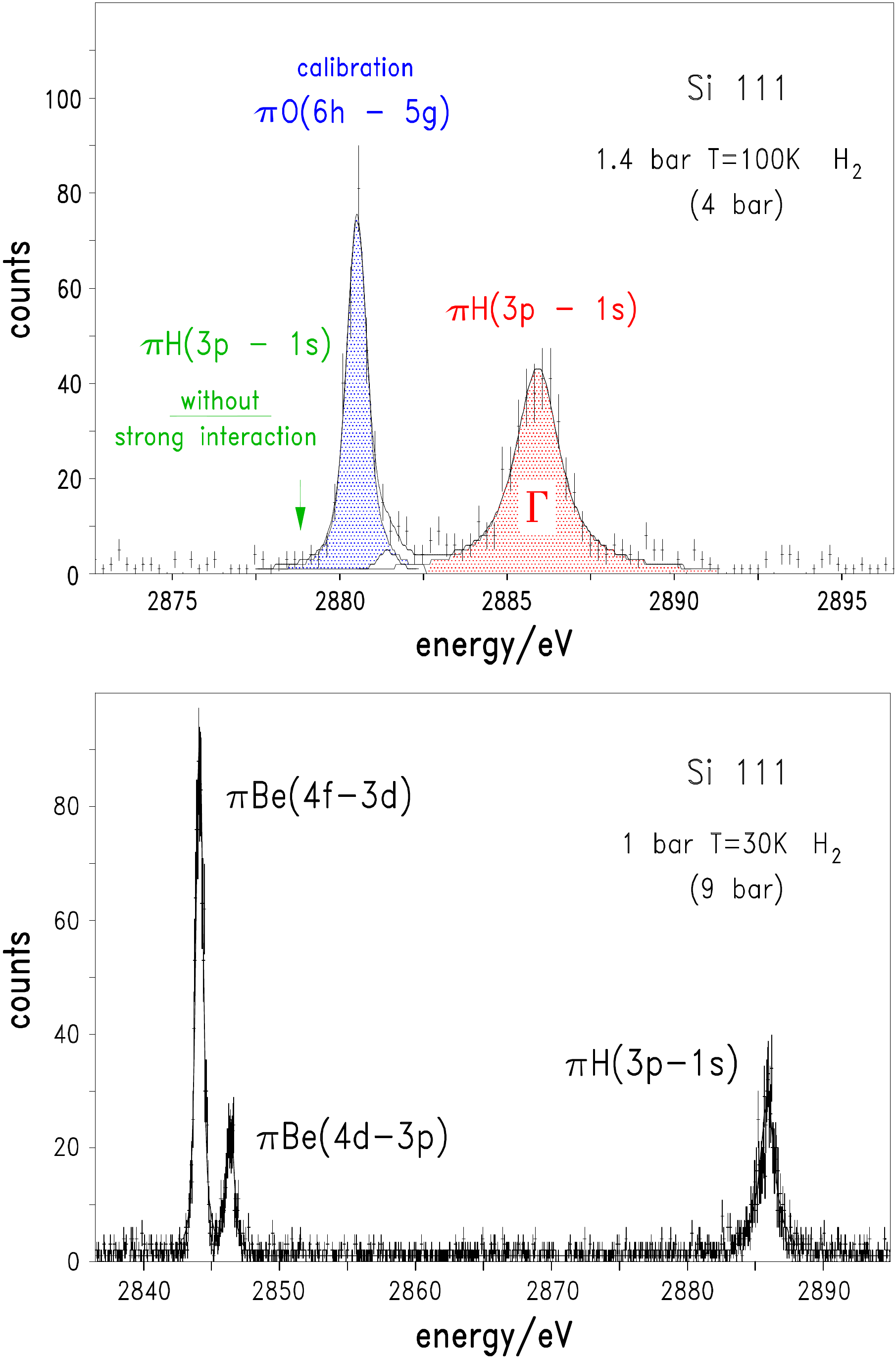}}
\caption{Spectra from simultaneous measurements of the $\pi$H$(3p-1s)$ transition and the $\pi^{16}$O$(6h-5g)$ (top) and the $\pi^{16}$Be$(4f-3d)$ calibration lines (bottom).}
\label{fig:spectra_1}
\end{figure}

\begin{figure}[]
\resizebox{0.42\textwidth}{!}{\includegraphics{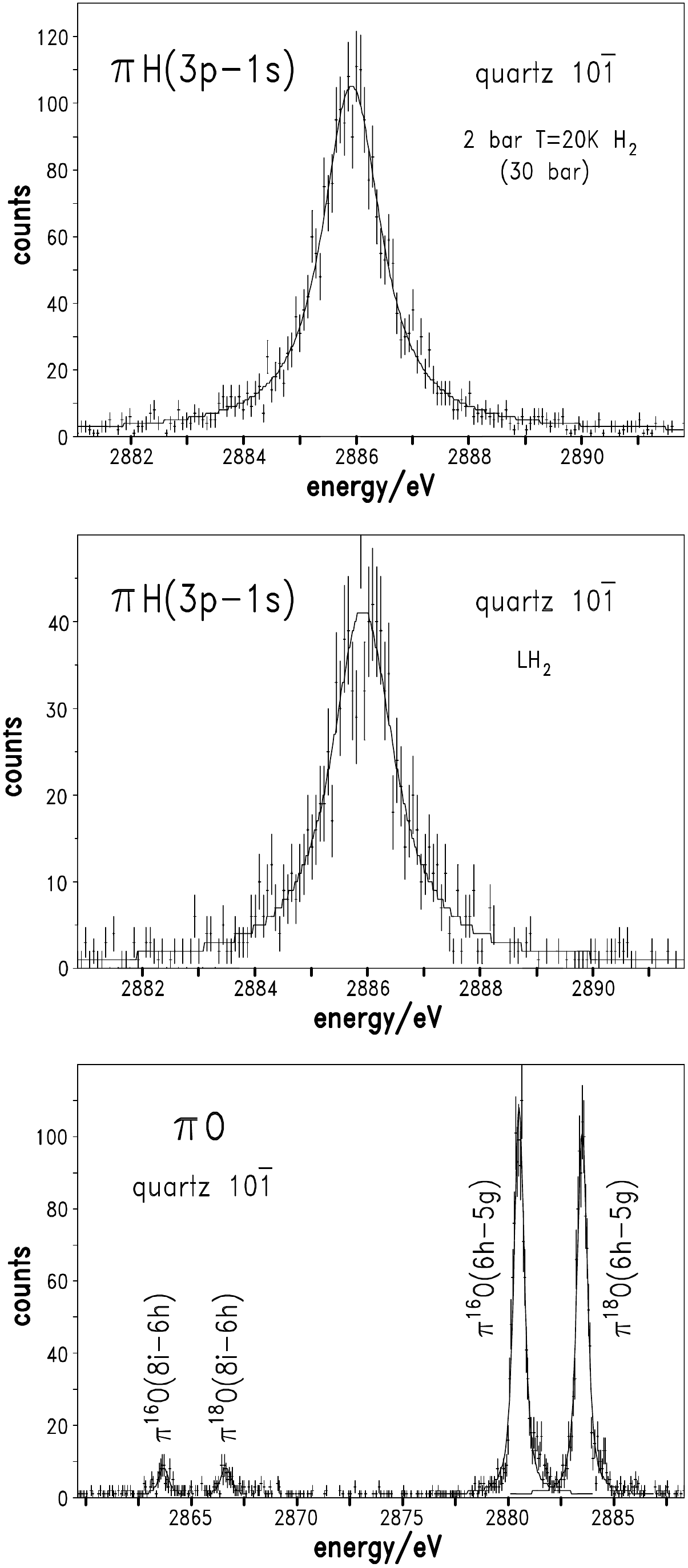}}
\caption{High statistics measurement of the $\pi$H$(3p-1s)$ transition (top),  first crystal spectrometer measurement  in liquid hydrogen (middle), and energy calibration with $(6-5)$ transitions from an $^{16}$O$_{2}/^{18}$O$_{2}$ mix target at 4\,bar equivalent density with an 80\% He admixture. Also visible are the inner transitions $(8i-6h)$ of the oxygen isotopes having energies of 2863.689 and 2866.652\,eV.}
\label{fig:spectra_2}
\end{figure}

In the Johann-type setup, the  absolute Bragg angle $\Theta_{\mathrm{Bragg}}$ cannot be determined with sufficient precision. Therefore, calibration is performed relative to a line of known energy with $\Theta_{\mathrm{Bragg}}$ as close as possible to the $\pi$H case. Experimentally, this corresponds to the determination of the angle  difference as measured by means of the position difference on the detector (figs.\,\ref{fig:spectra_1} and \ref{fig:spectra_2}).

In this experimental approach, the energy calibration is performed in two ways: 

(i) Fluorescence X-rays are well suited for stability monitoring because high rates are available when illuminating suitable materials  with commercial X-ray tubes. Here, the Zn K$\alpha_1$ and Ga K$\alpha_2$ lines are used having almost the same Bragg angle as the $\pi$H$(3p-1s)$ and the $\pi$H$(4p-1s)$ transition, respectively. However, the Zn and Ga lines have to be  measured in third order, whereas the $\pi$H transitions are reflected in first order. This requires a substantial correction due to the change of the index of reflection ($\bigtriangleup\Theta_{\mathrm{IRS}}$) (table\,\ref{table:Bragg_angles}). The calibration energy is given by the peak of the fluorescence line, which corresponds to the tabulated value. For Zn, the value was taken from the compilation of Deslattes et al.\,\cite{Des03}. Energies communicated for Ga X-rays were obtained from the compound GaAs as used in this experiment\,\cite{Moo07}. However, the precision achievable using fluorescence X-rays is limited due to the large natural width and the line asymmetry because of satellite structures\,\cite{Ana99}. Hence, these calibrations are mainly regarded as consistency checks.

(ii) For the first time, $\pi$H transitions were calibrated using X-rays from pionic atoms itself ($\pi$O$(6h-5g)$ and $\pi$Be$(4d-3p)$) (table\,\ref{table:exp_conditions} and fig.\,\ref{fig:spectra_1}) having natural line widths much smaller than Zn or Ga K X-rays. Furthermore, the uncertainty of the pion mass (2.5ppm) \cite{Ber12} cancels in leading order (see table\,\ref{table:results_R98}). Important, however, is the use of transitions from the intermediate part of the atomic cascade between circular orbits like $6h-5g$ or $4f-3d$, where level  energies are not affected by nuclear finite size and, therefore, do not experience a noticeable hadronic shift. Unfortunately, suitable pionic-atom transitions close in energy only   exist for the $\pi$H$(3p-1s)$ transition.

\setlength{\tabcolsep}{3.3mm}

\begin{table*}[t]
\begin{center}
\caption{Energies $E_\mathrm{X}$ and Bragg angles $\Theta_{\mathrm{Bragg}}$ for the various combinations of pionic hydrogen and calibration lines. Energies of fluorescence X-rays are taken from\,\cite{Des03}. Bragg angles for $\pi$H lines are calculated from the final result. The lattice constants are taken to be $2d=(0.6685977\,\pm\,0.000022)$\,nm for (natural) quartz\,\cite{Bri80} and $2d=(0.62712016\,\pm\,0.0000001)$\,nm for silicon\,\cite{Bas94} at $T$\,=\,$25^{\circ}$C. 
$\bigtriangleup\Theta_{\mathrm{IRS}}$, $\bigtriangleup\Theta_{\mathrm{p}}$, $\bigtriangleup\Theta_{\mathrm{b}}$, and $\bigtriangleup\Theta_{\mathrm{def}}$ are the angular corrections (in seconds of arc) to be applied due to index of refraction, bending, penetration depth, and defocussing (see text).}
\label{table:Bragg_angles}
\vspace{-4mm}
\begin{tabular}{lclcccccc}
\hline\\[-2mm]
crystal & reflection & transition          &$E_\mathrm{X}$ / eV & $\Theta_{\mathrm{Bragg}}$    & $\bigtriangleup\Theta_{\mathrm{IRS}}$ & $\bigtriangleup\Theta_{\mathrm{p}}$& $\bigtriangleup\Theta_{\mathrm{b}}$ & $\bigtriangleup\Theta_{\mathrm{def}}$\\[1mm]           
\hline\\[-2mm]
quartz & $10\bar{1}$ & $\pi$H$(3p-1s)$                     && 39$^{\circ}$59'0.3"  & 28.2" & -0.11" & -1.62" & 0 \\
       &     & $\pi^{18}$O$(6h-5g)$ &2883.488\,$\pm$\,0.001 & 40$^{\circ}$~1'26.0" & 28.2" & -0.11" & -1.63" & 0.01"\\
       &     & $\pi^{16}$O$(6h-5g)$ &2880.506\,$\pm$\,0.001 & 40$^{\circ}$~4'25.4" & 28.3" & -0.11" & -1.63" & 0.03"\\
\hline\\[-2mm]
silicon &111 & $\pi$H$(3p-1s)$                             && 43$^{\circ}$14'26.6" & 24.2" & -0.09" & -4.16" & 0 \\
      &      & $\pi^{16}$O$(6h-5g)$ &2880.506\,$\pm$\,0.001 & 43$^{\circ}$20'31.2" & 24.3" & -0.09" & -4.18" & 0.04"  \\
       &     & $\pi$Be$(4f-3d)$     &~\,2844.064\,$\pm$\,0.001$^{a}$& 44$^{\circ}$~2'20.0" & 24.8" & -0.09" & -4.28" & 0.46"\\[1mm]
\hline\\[-2mm]

silicon & 111 & $\pi$H$(3p-1s)$                            && 43$^{\circ}$14'26.6" & 24.2" & -0.09" & -4.16" & 0 \\       
        & 333 & Zn K$\alpha_1$      &8638.904\,$\pm$\,0.073 & 43$^{\circ}$21'30.1" &  2.7" & -0.44" & -4.04" & 0.05"\\[1mm]\hline\\[-2mm]

silicon & 333 & Ga K$\alpha_2$      &9224.835\,$\pm$\,0.027 & 40$^{\circ}$~0'44,1" &  2.4" & -0.50" & -3.59" & 0\\
        & 111 & $\pi$H$(4p-1s)$                            && 40$^{\circ}$30'58.4" & 22.1" & -0.10" & -3.78" & 0.31" \\[0.5mm]
\hline\\[-2mm]
\multicolumn{9}{l}{$^{a}$ assumes two remaining K and one L electron}
\end{tabular}\\
\end{center}
\end{table*}

\setlength{\tabcolsep}{2.0mm}
\begin{table*}
\begin{center}
\caption{Experimental conditions for the various measurements of $\pi$H transitions and calibration lines (label A - G). Equivalent hydrogen densities are given as pressure values corresponding to a temperature of 20$^\circ$C (NTP).  Counts are results of fits to the line shape.}
\label{table:exp_conditions}
\vspace{-4mm}
\begin{tabular}{clccccccrcc}
\hline\\[-2mm] 
label &~~~~~~~~transitions & crystal && equivalent &&counts&& \multicolumn{3}{c}{counts per transition}   \\ 
&hydrogen / calibration & reflection && density H$_2$ &&$\pi$H&& \multicolumn{1}{c}{~~$\pi$H}&&\multicolumn{1}{c}{$\pi$O, $\pi$Be or X-ray } \\ 
&                          &        && / bar         && / hour  \\ 
\hline\\[-2mm] 
A & $\pi$H$(3p-1s)$/$\pi$O$(6h-5g)^{a}$ & Si 111/Si 111 && 4.1~$\pm$~0.1 &&12.2~$\pm$~0.8&& 673~$\pm$~45 && 461~$\pm$~93\\ 
B & $\pi$H$(3p-1s)$/$\pi$O$(6h-5g)^{a}$ & qu 10$\bar{1}$/qu 10$\bar{1}$ && 3.8~$\pm$~0.1 &&17.0~$\pm$~0.7&& 1004~$\pm$~38 && 729~$\pm$~30\\ 
C & $\pi$H$(3p-1s)$/$\pi$O$(6h-5g)$ & qu 10$\bar{1}$/qu 10$\bar{1}$ && 29.7~$\pm$~0.6 &&37.8~$\pm$~0.5&& 6838~$\pm$~92 && 6867~$\pm$~91\,~~\\ 
D & $\pi$H$(3p-1s)$/$\pi$O$(6h-5g)$ & qu 10$\bar{1}$/qu 10$\bar{1}$ && LH$_2$ &&16.2~$\pm$~0.5&& 1719~$\pm$~49 && 693~$\pm$~29\\ 
E & $\pi$H$(3p-1s)$/$\pi$Be$(4f-3d)^{a}$ & Si 111/Si 111 && 9.3~$\pm$~0.3  &&22.1~$\pm$~0.9&& 928~$\pm$~37 && 1273~$\pm$~39~~~\\ 
F & $\pi$H$(3p-1s)$/Zn K$\alpha_1$ & Si 111/Si 333 && 9.5~$\pm$~0.3  &&40.0~$\pm$~0.5&& 7220~$\pm$~96 && 485380~$\pm$~909\,~~~~\\ 
G & $\pi$H$(4p-1s)$/Ga K$\alpha_2$ & Si 111/Si 333 && 9.9~$\pm$~0.3  &&14.7~$\pm$~1.2&& 265~$\pm$~22 && 2626~$\pm$~59~~~\\[0.5mm] 
\hline\\[-2mm] 
\multicolumn{6}{l}{$^{a}$ simultaneous measurement of $\pi$H transition and calibration line}
\end{tabular} 
\end{center}
\end{table*}

For gaseous targets, pionic oxygen can be regarded as a hydrogen-like atomic system, because screening effects from remaining electrons are small due to a practically complete removal of the electron shells before the\linebreak
$\pi$O$(6h-5g)$ transition occurs\,\cite{Bac89,Kir99,Ana03}. Line broadening owing to Coulomb explosion\,\cite{Sie00} is symmetric and, hence, does not affect the position. Similarly, Doppler broadening caused by Coulomb de-excitation\,\cite{Bra78,Bad01} can be assumed to be symmetric as well (see, \textit{e.\,g.}, ref.\,\cite{Jen03}).

At low density, $\pi$O$(6-5)$ and $\pi$H$(3p-1s)$ lines were measured simultaneously by using a hydrogen-oxygen gas mixture of H$_{2}$/O$_{2}$ (98\%/2\%) (fig.\,\ref{fig:spectra_1}). Higher densities\linebreak  
($\geq$\,10\,bar equivalent) require lower temperatures. Hence, hydrogen and oxygen were measured alternately because the oxygen gas freezes out in such conditions. In one of these measurements, a mixture of $^{4}$He/$^{16}$O$_{2}$/$^{18}$O$_{2}$ \linebreak
(80\%/10\%/10\%) was used (fig.\,\ref{fig:spectra_2}). The oxygen isotopes serve as a check for the dispersion of the spectrometer. The admixture of helium reduces self absorption of X-rays in the oxygen target.

For solid targets, like Be as used here, in contrast to gases, electron screening has been found to be significant\,\cite{Bac85,Jec94,Len98}. Here, a fast electron refilling is expected from the environment restoring electronic shells even during the rapidly proceeding atomic cascade. Moreover, for this mixed target setup H$_2$/Be the rate is limited because of the predominant absorption in the beryllium plates.

The limiting systematic uncertainty is due to  the simultaneous description of the parallel transitions $\pi$O$(6g-5f)$ or $\pi$Be$(4d-3p)$ contributing to the line shape on the high-energy side. The accuracy in the calculation of the pure electromagnetic transition energy itself is about one meV only\,\cite{Sch11}.

\section{Analysis}

\subsection{Data processing}

As discussed in detail in ref.\,\cite{Str11}, the Bragg reflection of a narrow X-ray line generates a parabolic hit pattern in the detector plane. Analyzing the hit pattern and acceptance of only single or two-pixel events significantly reduces  beam--induced background leading to an excellent peak-to-background ratio (figs.\,\ref{fig:spectra_1} and \ref{fig:spectra_2}). The curvature of the parabola was fitted to the narrow $\pi$O or the high-statistics X-ray fluorescence line and then applied to the corresponding $\pi$H transition. The one-dimensional spectrum obtained after projection of the curvature-corrected pattern to the axis of dispersion is equivalent to an energy spectrum.

The position of the reflection is obtained by means of a fit to the line shape using Voigt profiles. Though the spectrometer response is slightly asymmetric, it turned out that using a symmetric profile does affect the lines' position difference by typically 1-2\,meV only. For the determination of the hadronic broadening, however, the inclusion of the exact response is mandatory\,\cite{Got08,Cov09,Str10,Str11}.

Owing to imaging and diffraction properties, corrections must be applied to the line position as obtained from the fit (table\,\ref{table:Bragg_angles}). Main correction is the above-mentioned index of refraction shift $\bigtriangleup\Theta_{\mathrm{IRS}}$, in particular, when $\pi$H and calibration line are measured in different orders of reflection. Secondly, as  penetration depth of the X-radiation differs significantly with energy a corresponding correction has to be applied ($\bigtriangleup\Theta_{\mathrm{p}}$). Hence, an additional bending correction is necessary which takes into account the variation of the lattice constant with penetration depth ($\bigtriangleup\Theta_{\mathrm{b}}$). We use the approach of Cembali et al.\,\cite{Cem92}.

A further correction stems from the different focal
\linebreak 
lengths of $\pi$H and calibration lines ($\bigtriangleup\Theta_{\mathrm{def}}$). The shift due to this defocussing is approximately linear with the difference in the distance crystal-to-detector. 

The uncertainty of $\bigtriangleup\Theta_{\mathrm{IRS}}$ is on the order of 0.5\%\,\cite{Hen93,Cha95a}. A conservative estimate for the accuracy of $\bigtriangleup\Theta_{\mathrm{p}}$ and $\bigtriangleup\Theta_{\mathrm{b}}$ is 30\% which owes to deviations from the ideal crystal structure in the thin reflecting surface layer. However, for measurements using the same order of reflection only differences of rather small corrections contribute. Therefore, a maximum uncertainty of about 50\% for the difference can be assumed. The defocussing correction $\bigtriangleup\Theta_{\mathrm{def}}$ has been studied by means of Monte-Carlo simulation to be precise at least on the level of 20\%.

Further possible sources of uncertainties are the accuracy of the curvature correction, the distance of the Bragg crystal to the detector and its alignment and pixel size, as well as temperature stability. The analysis of position monitoring and a survey measurement of the experimental setup yields individual contributions of less than 0.2\,meV each. Uncertainties owing to the individual measurements are included in the corresponding systematic error as given in table\,\ref{table:results_R98} and are, in particular, due to fit model and long term stability.

\subsection{Electromagnetic transition energies}

For pionic hydrogen, the pure electromagnetic transition energies have been recalculated recently\,\cite{Sch11}. Partly, the results deviate from previous calculations by more than the experimental accuracy (see tables\,\ref{table:QED} and \ref{table:results_comparison}). Additional effects arising from strong interaction and polarisability are discussed below.

The influence of the strong interaction on the $3p\,$-level energies may be estimated from the sum of the angular-momentum averaged isoscalar and isovector $\pi N$ $p\,$-wave scattering volumes $c_{0}+c_{1}$. Numerical values can be found, {\it e.\,g.}, in ref.\,\cite{Koc86}. Applying the Trueman formula as given by\,\cite{Lam69,Lam70} results in negligibly small values of 7.4 and 5.8\,$\mathrm{\mu}$eV for the $3p$ and $4p$ state, respectively.

Effects due to the electric and hadronic  polarisabilities of proton and pion have been discussed in ref.\,\cite{Sch11}. The energy shift was found to be $(-0.6\,\pm\,0.6)$\,meV. Therefore, this contribution is neglected, too. 

\setlength{\tabcolsep}{0.7mm}
\begin{table}[h] 
\caption{Calculated values for the pure electromagnetic transition energies $E_{\mathrm{QED}}$ in $\pi$H (from\,\cite{Sch11}). Values include the energy decrease due to atom recoil.}
\label{table:QED} 
\begin{tabular}{cccccccccccccc} 
\hline\\[-2mm]
transition &&& \multicolumn{3}{c}{$\pi$H$(3p-1s)$} &&&& \multicolumn{3}{c}{$\pi$H$(4p-1s)$}&\\
\hline\\[-2mm]
$E_{\mathrm{QED}}$ / eV &~~&& 2878.8303 &$\pm$&0.0064 &&~~&& 3036.0921 &$\pm$& 0.0068& \\[0.5mm] 
\hline\\
\end{tabular} 
\end{table}

\section{Results}

\subsection{Hadronic shift}

The results obtained from the individual measurements are consistent within the errors (table\,\ref{table:results_R98}). Therefore, a\linebreak weighted average is calculated: 
\begin{equation}
\epsilon_{1s} = 7.0858\,\pm\,0.0071\,(stat)\,\pm\,0.0064\,(sys)~\mathrm{eV}\,.
\label{eq:eps1s}
\end{equation}

The first error represents the statistical accuracy. The second one includes all systematic effects, which are due to the spectrometer setup, imaging properties of extended Bragg crystals, analysis and instabilities as well as the pion mass. The contribution from the uncertainty of the pion mass cancels in leading order for transitions calibrated with a pionic-atom transition itself (table\,\ref{table:results_R98}). 

The combined result of the most recent previous precision measurements yielded $\epsilon_{1s}=7.108\pm 0.036$\,eV\,\cite{Sch99,Sch01}, where the energy calibration was performed with argon K$\alpha$ fluorescence X-rays (the sign convention in refs.\,\cite{Sch99,Sch01} is opposite to the one used here). The measured\linebreak
$\pi$H$(3p-1s)$ transition energies, however, are in perfect agreement (table\,\ref{table:results_comparison}). Therefore, any discrepancy to the new result vanishes completely when taking  into account the newly calculated pure electromagnetic transition energy.

\setlength{\tabcolsep}{2.4 mm}

\begin{table*}[t]
\begin{center}
\caption{Results from this experiment contributing to the weighted average for the ground-state shift $\epsilon_{1s}$ in pionic hydrogen as given in eq.\,(\ref{eq:eps1s}) and table\,\ref{table:results_comparison}. $\bigtriangleup m_{\pi}$ denotes the contribution arising from the uncertainty of the pion mass to the error of line energy $E_\mathrm{exp}$ or hadronic shift $\epsilon_{1s}$.}
\label{table:results_R98}
\vspace{-4mm}
\begin{tabular}{clcccccccccc}
\hline\\[-2mm] 
label &~~~~~~~~transitions && $E_\mathrm{exp}$   & \multicolumn{4}{c}{$\bigtriangleup E_\mathrm{exp}$} && $\epsilon_{1s}$ & \multicolumn{2}{c}{$\bigtriangleup\epsilon_{1s}$}  \\ 
      & hydrogen / calibration &   &&{\it stat} & {\it sys exp} & QED & $\bigtriangleup m_{\pi}$ &&   &  $\bigtriangleup m_{\pi}$  & {\it total} \\ 
 &  &&      &      &       & ~or $E_\mathrm{X}$ &  &&  & &\\ 
 &  && / eV & / eV & / eV  & / eV      & / eV &&/ meV & / meV & / meV \\ 
\hline\\[-2mm] 
A & $\pi$H$(3p-1s)$/$\pi$O$(6h-5g)$  && 2885.8777 & 0.0439 & 0.0033 & 0.0014 & 0.0063 && 7047.4& 0.8& 44.1\\ 

B & $\pi$H$(3p-1s)$/$\pi$O$(6h-5g)$  && 2885.9010 & 0.0314 & 0.0036 & 0.0014 & 0.0063 && 7070.7& 0.8& 31.6\\ 

C & $\pi$H$(3p-1s)$/$\pi$O$(6h-5g)$  && 2885.9205 & 0.0120 & 0.0055 & 0.0014 & 0.0063 && 7090.2& 0.8& 13.6\\ 

D & $\pi$H$(3p-1s)$/$\pi$O$(6h-5g)$  && 2885.9160 & 0.0256 & 0.0055 & 0.0014 & 0.0063 && 7085.7& 0.8& 26.2\\ 

E & $\pi$H$(3p-1s)$/$\pi$Be$(4f-3d)$ && 2885.9207 & 0.0345 & 0.0045 & 0.0014 & 0.0063 && 7090.6& 0.8& 34.8\\ 

F & $\pi$H$(3p-1s)$/Zn K$\alpha_1$   && 2885.9280 & 0.0108 & 0.0125 & 0.073 & 0.0067 && 7097.7& 6.7& 75.1\\ 

G & $\pi$H$(4p-1s)$/Ga K$\alpha_2$   && 3043.1971 & 0.0598 & 0.0097 & 0.027 & 0.0067 && 7105.0& 6.7& 66.7\\[0.5mm] 
\hline 
\end{tabular} 
\end{center}
\end{table*}

\setlength{\tabcolsep}{3.3mm}

\begin{table*}
\begin{center}
\caption{Comparison of results for the ground-state shift $\epsilon_{1s}$ in pionic hydrogen. Values for $\epsilon_{1s}$ of previous experiments are given based on the corresponding values for the electromagnetic transition energy $E_\mathrm{QED}$, which are different from the most recent result of Schlesser et al.\,\cite{Sch11} as given in the last row. Noteworthy, that all precision experiments are in good agreement when using the same value for the electromagnetic transition energy (see text).}
\label{table:results_comparison}
\vspace{-2mm}
\begin{tabular}{lcccclrc}
\hline\\[-2mm] 
transition & equivalent & energy      & \multicolumn{1}{c}{~~~~~$E_\mathrm{exp}$}  & \multicolumn{1}{c}{~~~~$E_\mathrm{QED}$} & \multicolumn{1}{c}{~~$\epsilon_{1s}$} &&  \\ 
                & density H$_2$& calibration &  &  &  &  &    \\ 
                & / bar  &             & \multicolumn{1}{c}{~~~~~/ eV} & \multicolumn{1}{c}{~~~~/ eV} & \multicolumn{1}{c}{~/ meV} &&   \\ 
\hline\\[-2mm] 
$\pi$H$(2p-1s)$ & 4 & $^{55}$Fe/$^{65}$Zn& ~~~2500~$\pm$~500 & \multicolumn{1}{c}{~~~~\,2436} & \multicolumn{1}{c}{~~---} && \cite{Bai70}\\ 

$\pi$H$(2p-1s)$ & 18 & V K$\alpha_1$ & 2433.5~$\pm$~1.7 & 2429.6~$\pm$~0.2 & ~~\,3900~$\pm$~1700 && \cite{Fos83} \\

$\pi$H$(2p-1s)$ & 40 & V K$\alpha_1$ & 2434.5~$\pm$~0.5 & \multicolumn{1}{c}{~~~~~2429.6} & ~~\,4900~$\pm$~500 && \cite{Bov85} \\ 

$\pi$H$(3p-1s)$ & 15 & Ar K$\alpha_1$ & 2885.98~$\pm$~0.23 & 2878.86~$\pm$~0.03 & ~~\,7120~$\pm$~320 && \cite{Bee91}\\

$\pi$H$(3p-1s)$ & 15 & Ar K$\alpha_{1,2}$  & 2885.935~$\pm$~0.048 & 2878.808~$\pm$~0.008 & ~~\,7127~$\pm$~46 && \cite{Sig96a,Sig95} \\ 

$\pi$H$(4p-1s)$ & 15 & Ga K$\alpha_{2}$  & 3042.97~$\pm$~0.15 & 3036.068~$\pm$~0.009 & ~~\,6900~$\pm$~150 && \cite{Sig96a,Sig95}\\

$\pi$H$(3p-1s)$ & 15 & Ar K$\alpha_{1,2}$  & 2885.916~$\pm$~0.034 & 2878.808~$\pm$~0.008 & ~~\,7108~$\pm$~36 && \cite{Sch99,Sch01} \\[2mm] 

\multicolumn{5}{l}{\textit{weigthed average of this experiment (for details see table\,\ref{table:results_R98})}} & 7085.8~$\pm$~9.6 & \\[1mm] 
\hline 
\end{tabular} 
\end{center}
\end{table*}

\begin{figure}[]
\resizebox{0.51\textwidth}{!}{\includegraphics{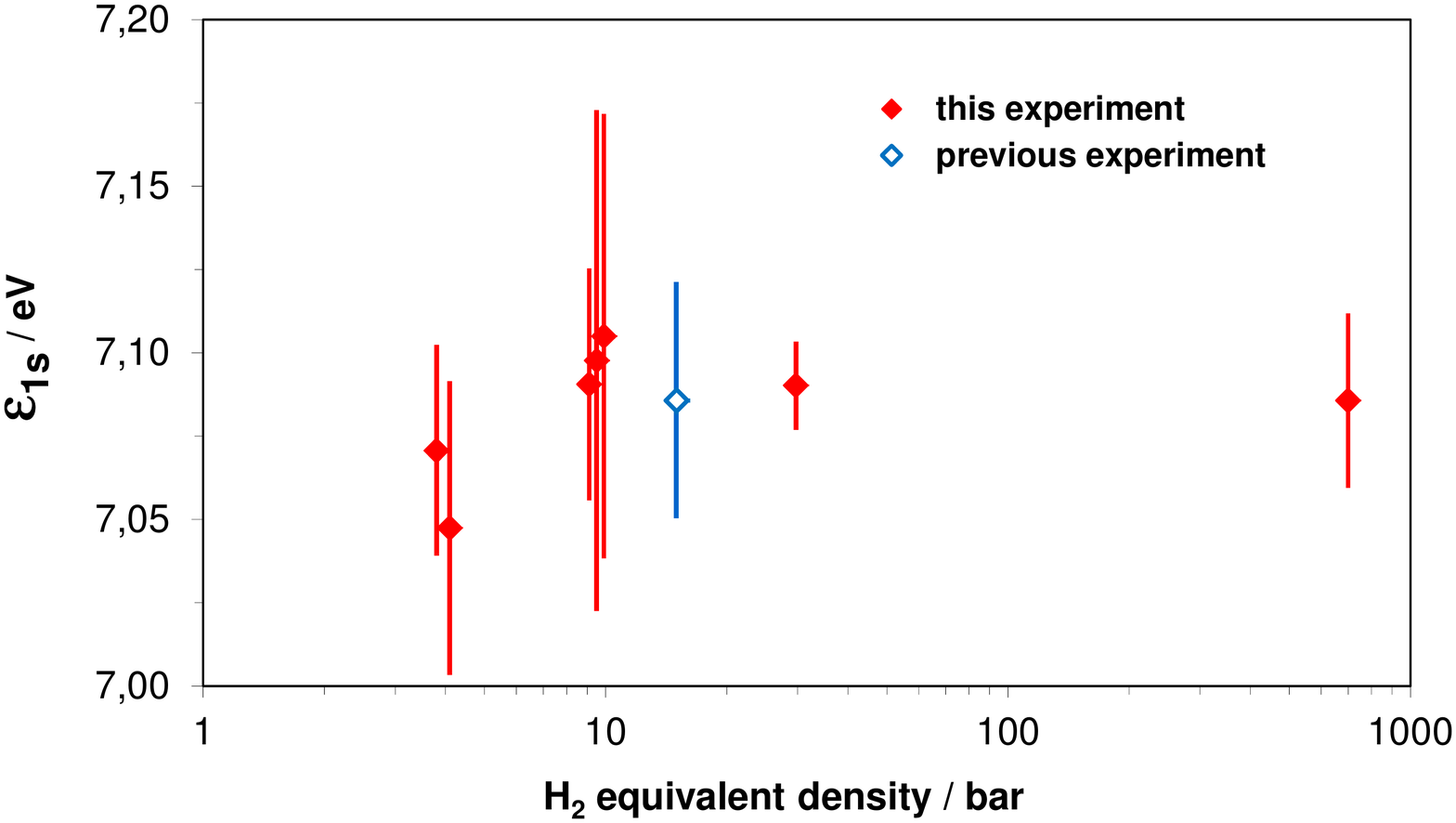}}
\vspace{-8mm}
\caption{Results for the hadronic ground-state shift $\epsilon_{1s}$ in pionic hydrogen at various H$_2$ densities 
($\blacklozenge$). The open diamond ($\diamondsuit$) represents the weigthed average of the previous precision experiment\,\cite{Sch01} when also using the new value for the electromagnetic transition energy. Note the logarithmic scale for the density.}
\label{fig:eps_density}
\end{figure}

As discussed in section\,\ref{sec:cascade}, a density dependence of the $\pi$H$(3p-1s)$ transition energy would be evidence for radiative decay after molecular formation. In this experiment, no evidence is found within the experimental uncertainty  between 3.8\,bar and liquid (fig.\,\ref{fig:eps_density}), {\it i.\,e.}, within a density range of a factor of about 200. A linear fit yields for the slope a value consistent with zero (($0.004\pm 0.04$)\,meV/bar). Consequently, the result for $\epsilon_{1s}$ is unaltered when extra\-polating to zero density. 

Remarkably, the result for the combination\linebreak
$\pi$H$(3p-1s)$/$\pi$Be$(4f-3d)$ is only consistent with the other measurements when the presence of two K electrons in the $\pi$Be system is assumed. The screening correction is calculated to be --\,143\,meV, which is about a factor of 4 larger than the experimental error for the transition energy. A complete refilling of electron shells is expected because capture rates are very high in particular from the conduction band in solid state metals\,\cite{Bac85}. The influence of an additional L electron screening (--\,9\,meV) is beyond the experimental accuracy.

\subsection{Scattering length}

Using the result given in eq.\,(\ref{eq:eps1s}), the Coulomb corrected  strong-interaction $\pi^-p$ scattering length is obtained from eq.\,(\ref{eq:eps_moddeser}) to be
\begin{equation}
a_{\pi^-p} = (85.26\,\pm\,0.12)\cdot 10^{-3}\,m^{-1}_{\pi},\label{eq:a_pi-p}
\end{equation}
which is 1.01\% smaller than the LO value when using eq.\,(\ref{eq:deser}).

The experimental uncertainty for $a_{\pi^-p}$  amounts to\linebreak  0.14\% only. However, applying eq.\,(\ref{eq:api-p_ChiPT}) for the extraction of the QCD quantity $a^{+}+a^{-}$ yields 
\begin{equation}
a^{+}+a^{-} = (81.4\,\pm\,3.6)\cdot 10^{-3}\,m^{-1}_{\pi}.\label{eq:a_a++a-}
\end{equation}
The uncertainty is due to the poor knowledge of low-energy constants appearing already in NLO as discussed in detail in refs.\,\cite{Bar11a,Bar11b}. The new value for $\epsilon_{1s}$ yields a small shift for the isoscalar scattering length from $a^+=(7.6\,\pm\,3.1)\cdot 10^{-3}\,m^{-1}_{\pi}$ to $(7.5\,\pm\,3.1)\cdot 10^{-3}\,m^{-1}_{\pi}$ in the analysis of Baru et al.\,\cite{Bar11a,Bar11b}.

\section{Summary}

The $\pi^-p$ scattering length has been determined with a precision of 0.14\% from various ground-state transitions in pionic hydrogen and applying different energy calibration schemes. The result represents a fourfold improvement compared to the most precise previous experiment\,\cite{Sch01}. All individual results coincide very well within the experimental uncertainties. The weighted average is in excellent agreement with the previous precision experiment  when the updated result for the pure electromagnetic transition energy is applied. 

Within the experimental uncertainty, no pressure dependence of the $\pi$H$(3p-1s)$ energy could be observed. This excludes, in particular, radiative decay out of molecular bound states, the formation of which is assumed to be density dependent. Any contribution above the 1\% level can be discarded. 

It is worthwhile to mention that the assumption is confirmed that exotic atoms formed within metallic targets recover electrons after Auger emission on time scales much faster than the cascade time in contrast to gaseous targets. This is revealed by the energy measured for the $\pi$Be ($4f-3d$) transition, which requires the presence of a complete electronic K shell at the time of the pionic transition.

\section*{Acknowledgements}

We would like to thank C.\,Hanhart and M.\,Hoferichter for discussions on various issues related to scattering lengths. We are grateful to N.\,Dolfus, H.\,Labus, B.\,Leoni and \linebreak K.-P.\,Wieder for solving numerous technical problems as well as N.\,Nelms for the help in X-ray detector issues. Efforts are acknowledged of the PSI staff to provide excellent beam conditions and by the Carl Zeiss AG, Oberkochen, Germany, in the fabrication of the Bragg crystals. We thank Prof. Dr. E.\,F\"{o}rster and his collaborators at the University of Jena and A.\,Freund and his group at ESRF for the help in characterising the crystal material. We are indebted to PSI for supporting the stay during the run periods (D.\,F.\,A. and Y.-W.\,L.). Partial funding was granted by the Allianz Program of the Helmholtz Association contract no. EMMI HA-216 "Extremes of Density and Temperature: Cosmic Matter in the Laboratory" (P.\,I. and M.\,T.). This work is part of the PhD thesis of one of us (M.\,H., Universit\"at zu K\"oln, 2003).

\end{document}